# Signatures of symmetry-stabilized Dirac fermions in correlated oxide $CaNbO_3$ thin films

Yunkyu Park[1+], Seoung-Hun Kang[2,3+], Jeongkeun Song[1+], Minsik Kong[4], Seohee Kim[4], Shan Lin[1], Eun Sang Choi[5], Km Rubi[6], Hua Zhou[7], Sangmoon Yoon[8], Hu Miao[1], Satoshi Okamoto[1], Jong Mok Ok[4], and Ho Nyung Lee[1*]

[1]Materials Science and Technology Division, Oak Ridge National Laboratory, Oak Ridge, TN 37831, USA

[2]Department of Physics, Kyung Hee University, Seoul, 02447, Republic of Korea

[3]Department of Information Display, Kyung Hee University, Seoul, 02447, Republic of Korea

[4]Department of Physics, Pusan National University, Busan 46241, Republic of Korea

[5]National High Magnetic Field Laboratory, Florida State University, Tallahassee, FL, 32310, USA

[6]National High Magnetic Field Laboratory, Los Alamos National Laboratory, Los Alamos, NM, 87544, USA

[7]Advanced Photon Source, Argonne National Laboratory, Lemont, IL, 60439, USA

[8]Department of Semiconductor, Gachon University, Seongnam 13120, Republic of Korea

[+] These authors contributed equally to this work. * Correspondence to: hnlee@ornl.gov

## Abstract

Symmetry-protected electronic states in correlated oxides offer a route to combining topology, lattice tunability, and many-body interactions, but creating such topological electronic states in complex oxides remains challenging. Here, we show that oxygen octahedral rotations provide a symmetry-control parameter for stabilizing Dirac-related electronic states in epitaxial $CaNbO_3$ films. By comparing compressively strained tetragonal films and relaxed orthorhombic films, we demonstrate that distinct octahedral rotation patterns preserve nonsymmorphic symmetry and lead to distinct transport signatures consistent with Dirac-like electronic states. The films exhibit metallic transport with high mobility up to $\mu \approx 138{,}000\ cm^2 \cdot V^{-1} \cdot s^{-1}$, multichannel conduction, and Shubnikov–de Haas oscillations from small high-mobility Fermi pockets. Landau-fan diagram analysis indicates finite Berry-phase contributions, while first-principles calculations reveal a strain-driven evolution from nodal-line band crossings in the orthorhombic phase to Dirac-electron band features in the tetragonal phase. These results establish oxygen octahedral symmetry as a design knob for engineering topological electronic states in correlated perovskite oxides.

## Introduction

The interplay between electronic correlations and topological band structures offers a promising route toward emergent quantum phenomena in solids.[1-5] Transition-metal oxides are particularly attractive because their lattice, orbital, spin, and charge degrees of freedom are strongly intertwined and can be continuously tailored through epitaxial engineering[2]. Unlike weakly correlated topological semimetals,[6-8] oxide systems exhibit intimate coupling between lattice distortions and electronic structure, providing a unique opportunity to control symmetry-derived electronic states through structural design. Despite such potential, realizing symmetry-protected Dirac states in correlated oxides remains challenging because strong structural distortions often obscure or destroy band degeneracies near the Fermi level.[9-14]

Recent studies[15-18] have identified nonsymmorphic crystal symmetry as a powerful mechanism for stabilizing Dirac and nodal electronic states in oxide perovskites. In these systems, oxygen octahedral rotations can generate glide or screw symmetries through fractional lattice translations (see Figs. 1a and 1b), enabling symmetry-protected band crossings at the Brillouin-zone boundary.[15-19] Because octahedral rotations are highly tunable by epitaxial strain,[20-23] they provide a unique structural pathway for engineering topological electronic states in correlated oxides. Yet, whether distinct octahedral rotation patterns can preserve nonsymmorphic symmetry across different structural phases and sustain Dirac-related electronic states remains largely unexplored.

Among oxide Dirac-semimetal candidates, 4*d* niobates have emerged as a promising platform because moderate electron correlations coexist with relatively dispersive Nb 4*d* bands.[24,25] Previous studies on $SrNbO_3$ demonstrated that epitaxial strain can induce nonsymmorphic symmetry and Dirac-electron transport behavior through a cubic-to-tetragonal structural transition.[17,19,21-23,26,27] However, the higher structural symmetry of $SrNbO_3$ limits the accessible octahedral rotation manifold and therefore restricts the range of symmetry engineering. In contrast, $CaNbO_3$ intrinsically hosts lower symmetry orthorhombic distortions with multiple competing octahedral rotation patterns, providing a broader structural platform for investigating how nonsymmorphic symmetry evolves across distinct lattice phases.[28]

Here, we demonstrate that oxygen octahedral rotations stabilize symmetry-derived Dirac-related electronic states in epitaxial $CaNbO_3$ films. By comparing compressively strained tetragonal films and relaxed orthorhombic films, we show that distinct octahedral rotation patterns preserve nonsymmorphic crystal symmetry across structural phases. Both phases exhibit high carrier mobility and Shubnikov–de Haas (SdH) oscillations arising from small Fermi pockets with finite Berry-phase contributions. These experimentally observed Fermi-surface characteristics are well captured by first-principles calculations, which predict a strain-driven evolution from Dirac-like band crossings in the tetragonal phase to nodal-line-like electronic structures in the orthorhombic phase. These results establish oxygen octahedral symmetry as a versatile design parameter for engineering symmetry-derived topological phenomena in correlated oxides.

## Results

To establish how oxygen octahedral rotations (OORs) govern nonsymmorphic symmetry in $CaNbO_3$, we first examine the strain-dependent evolution of octahedral rotation patterns. Unlike $SrNbO_3$, where nonsymmorphic symmetry emerges only after a cubic-to-tetragonal transition, $CaNbO_3$ intrinsically hosts structurally distorted phases that preserve symmetry-derived band degeneracies across multiple OOR configurations. In the orthorhombic phase, several distinct

OOR patterns are symmetry-allowed, including $a^+b^+c^+$ (space group: *Immm*), $a^+b^-b^-$ (space group: *Pnma*), $a^0b^+b^-$ (space group: *Cmcm*), and $a^0b^-b^-$ (space group: *Imma*), whereas the tetragonal phase stabilizes either $a^0a^0c^+$ (space group: *P4/mbm*) or $a^0a^0c^-$ (space group: *I4/mcm*) rotations. Based on our previous structural study[28], relaxed $CaNbO_3$ films adopt orthorhombic $a^+b^-b^-$/$a^0b^+b^-$ distortions, whereas compressively strained films stabilize the tetragonal $a^0a^0c^-$ phase (Fig. 1c). Unlike $SrNbO_3$, where the bulk phase is cubic and nonsymmorphic symmetry emerges only after strain-induced tetragonal distortion, both strained and relaxed $CaNbO_3$ intrinsically possess symmetry-lowering octahedral distortions.[28] These intrinsic distortions fundamentally redefine the symmetry landscape by preserving nonsymmorphic operations through distinct OOR patterns in both structural phases.

In the compressively strained tetragonal phase, the $a^0a^0c^-$ rotation pattern preserves a glide-like symmetry operation consisting of mirror reflection combined with fractional lattice translation. This symmetry constrains the electronic structure at the Brillouin-zone boundary and stabilizes isolated Dirac band crossings near the Fermi level, giving rise to small symmetry-derived Fermi pockets. In contrast, the lower-symmetry orthorhombic phase retains nonsymmorphic symmetry together with additional mirror and inversion constraints. Rather than producing isolated Dirac points, the combined symmetry protection stabilizes extended band degeneracies along high-symmetry directions, generating nodal-line-like electronic structures near the Fermi level. Consequently, the orthorhombic phase is expected to host more anisotropic multiband Fermi surfaces while preserving Dirac-related transport characteristics. Figure 1d summarizes the strain-driven evolution of OOR symmetry and the associated electronic topology in $CaNbO_3$. Accordingly, both phases are expected to exhibit Dirac-related transport signatures, including high carrier mobility[19,26,29,30] and nontrivial Berry phase.[31,32]

To experimentally probe these symmetry-derived transport properties, we synthesized both strained (4 nm in thickness) and strain-relaxed (40 nm in thickness) $CaNbO_3$ films on (001)-oriented $SrTiO_3$ substrates using pulsed laser deposition (PLD). Reflection high-energy electron diffraction (RHEED) patterns exhibit sharp streaks and well-defined diffraction spots, indicating layer-by-layer growth and high crystalline quality (Fig. 2a). High-angle annular dark-field scanning transmission electron microscopy imaging reveals atomically abrupt interfaces. X-ray diffraction confirms phase-pure epitaxial $CaNbO_3$ without secondary phases (Fig. 2b). Based on our previous structural analysis[28], 4 nm films stabilize the strained tetragonal $a^0a^0c^-$ phase, whereas 40 nm films relax into the orthorhombic $a^+b^-b^-$ phase (Figure S1). These samples are therefore referred to as strained tetragonal (ST) and relaxed orthorhombic (RO) $CaNbO_3$ films, respectively.

To probe the electrical transport properties of the ST- and RO-$CaNbO_3$ films, we measured the temperature-dependent resistivity of strained and relaxed $CaNbO_3$ films (Fig. 2c). Both films exhibit highly metallic behavior with exceptionally large residual resistivity ratios (RRR $\approx 10^3 - 10^4$), indicative of low disorder and high crystalline quality. The low-temperature resistivity follows a power-law dependence $\rho(T) = \rho_0 + A \cdot T^n$ with $n = 2.08$ below 50 K, consistent with dominant electron–electron scattering and Fermi-liquid transport behavior. To further investigate the transport characteristics, we performed magnetic-field-dependent Hall measurements. Both films exhibit nonlinear Hall resistivity (Figure S2), indicating multicarrier transport. The Hall response is well described by a semiclassical two-carrier model within Boltzmann transport theory.[33] From the fitting analysis, we extracted the temperature-dependent carrier densities and mobilities (Fig. 2d). At 2 K, the RO-$CaNbO_3$ film exhibits a carrier density $n_{3D} \approx 4.2 \times 10^{21}$ cm$^{-3}$ with a mobility $\mu \approx 22{,}800$ cm$^2 \cdot V^{-1} \cdot s^{-1}$, whereas the ST-$CaNbO_3$ film reaches an ultrahigh mobility of 138,000 cm² $V^{-1}$ $s^{-1}$ with a lower carrier density of $n_{3D} \approx 2.6 \times 10^{21}$ $cm^{-3}$. The substantially enhanced mobility in the strained phase suggests reduced carrier scattering arising from symmetry-derived Dirac electronic states near the Fermi level. These mobilities are comparable to those reported in representative oxide semimetals and high-mobility oxide heterostructures (Fig. S3). Note that the charge accumulation or the formation of a two-dimensional electron gas in $SrTiO_3$ can occur at oxide interfaces and may contribute to high-mobility transport[34-36].

To obtain insight into the transport behavior in ST- and RO-$CaNbO_3$ films, we next performed high-field magnetotransport measurements to probe their Fermi-surface properties. As shown in Fig. 3a-b, angle-dependent magnetoresistance measurements were carried out at 400 mK under magnetic fields up to 31 T with the field maintained perpendicular to the current direction (B ⊥ I). Both films exhibit clear SdH oscillations in the longitudinal resistivity over a wide angular range, indicating the formation of Landau quantized electronic states near the Fermi level (Fig.S*). The persistence of SdH oscillations across multiple field orientations indicates that the corresponding Fermi surfaces possess substantial three-dimensional rather than two-dimensional character [37,38]. This three-dimensional Fermi-surface character in both ST- and RO- $CaNbO_3$ supports the interpretation that the observed high-mobility transport and quantum oscillations predominantly originate from the intrinsic electronic states of $CaNbO_3$ rather than from a quasi two-dimensional electron gas in $SrTiO_3$[35,36].

Although both films exhibit three-dimensional Fermi surfaces, their quantum oscillation characteristics differ markedly. The ST-$CaNbO_3$ film displays relatively weak and broad SdH oscillations with a low oscillation frequency of $B_F$ = 4.41 T, corresponding to a Fermi surface cross-sectional area $S_F \approx 4.21 \times 10^{-4}$Å$^{-2}$ based on the Onsager relation. In contrast, the RO-$CaNbO_3$ film exhibits more pronounced oscillations with a higher frequency $B_F$ = 11.53 T, corresponding to $S_F \approx 1.10 \times 10^{-3}$Å$^{-2}$. From these Fermi surface areas, the carrier densities estimated from SdH oscillations are $n_{SdH} \approx 5.29 \times 10^{16}$ cm$^{-3}$ and $2.20 \times 10^{17}$ cm$^{-3}$ for the ST- and RO-$CaNbO_3$ films, respectively. These values are several orders of magnitude smaller than those obtained from Hall measurements, reflecting the multiband nature of $CaNbO_3$, where quantum oscillations selectively probe small high-mobility Fermi surface pockets. The observation of such low-frequency oscillations suggests that these pockets originate from symmetry-derived electronic states that are particularly sensitive to nonsymmorphic symmetry.

To further analyze the quantum oscillations, we extracted the oscillatory component by subtracting a smooth background from the longitudinal resistivity. Fig. 3c-d display the resulting oscillatory component ($\Delta\rho_{xx}$) as a function of inverse magnetic field, analyzed using the Lifshitz–Kosevich formalism. The fitting yields effective masses ($m^*$) and Dingle temperatures ($\tau_q$); the $m^* \approx 0.70\ m_0$ and $\tau_q \approx 3.0 \pm 0.3$ K for the ST-$CaNbO_3$ film and $m^* \approx 0.76\ m_0$ and $\tau_q \approx 1.95 \pm 0.15$ K for the RO-$CaNbO_3$ film, indicating relatively light carriers compared with those typically observed in correlated oxides[32,39-47]. For RO-$CaNbO_3$ film, consistent effective masses were independently obtained from the temperature dependence of the oscillation amplitude (Figure S4). Note that the reduced thickness of the ST-$CaNbO_3$ film makes it difficult to reliably extract the effective mass from the temperature dependence of the oscillation amplitude, since the oscillatory component is relatively weak. The corresponding quantum mobilities are $\mu_q \approx 1.0 \times 10^3$ cm$^2$/V·s for the ST-$CaNbO_3$ film and $\mu_q \approx 1.4 \times 10^3$ cm$^2$ V$^{-1}$·s$^{-1}$ for the RO-$CaNbO_3$ film, suggesting sufficiently long quantum lifetimes to sustain coherent cyclotron motion.

To further examine the nature of the charge carriers, we constructed Landau fan diagrams by assigning integer Landau indices ($N$) to the minima in $-d^2\rho_{xx}/dB^2$ (Fig. 3c-d). In the high-field regime where $\rho_{xy} >> \rho_{xx}$, minima in $\rho_{xx}$ approximately correspond to minima in the longitudinal conductivity $\sigma_{xx}$, allowing integer index assignment. The resulting Landau fan diagrams (Figure S5) exhibit clear linear behavior with the intercept of the Landau fan diagram ($\varphi_B/2\pi + \delta$) to be 0.28 for the ST-$CaNbO_3$ film and 0.57 for the RO-$CaNbO_3$ film. These nonzero intercepts are consistent with finite Berry-phase contributions associated with nontrivial electronic states.

The coexistence of light effective masses, high quantum mobility, and finite Berry-phase contributions suggests that the low-frequency oscillations originate from symmetry-derived electronic states. To gain further insight into the origin of the small Fermi pockets and finite Berry-phase signatures, we performed first-principles calculations for the strained tetragonal $a^0a^0c^-$ and relaxed orthorhombic $a^+b^-b^-$ phases of $CaNbO_3$. As shown in Fig. 4a, the strained tetragonal phase hosts Dirac band crossings near the Fermi level at the Brillouin-zone boundary, stabilized by nonsymmorphic symmetry associated with the out-of-phase oxygen octahedral rotation. In contrast, the orthorhombic phase found in the strain-relaxed $CaNbO_3$ exhibits extended band degeneracies along high-symmetry directions, forming nodal-line-like electronic structures protected by combined nonsymmorphic and mirror symmetries, as highlighted by the red dashed lines in Fig. 4b. These symmetry-derived band features provide an explanation for the finite Berry-phase contributions observed in both films.

We further analyzed the three-dimensional Fermi surfaces to establish their connection with the experimentally observed quantum oscillations. For direct comparison with the quantum oscillation measurements, we considered the $B||z$ configuration, where the cyclotron motion occurs within the $k_x$–$k_y$ plane. Consecutive $k_x$–$k_y$ slices were examined, and the enclosed areas of closed Fermi contours were converted into oscillation frequencies using the Onsager relation. For the strained tetragonal phase, the nominal Fermi level does not produce a small, closed orbit consistent with the experimentally observed SdH frequency of 4.41 T. However, an upward shift of the chemical potential by approximately + 220 meV produces a small hole-like pocket with a calculated frequency of approximately 4.37 T, in quantitative agreement with experiment. This agreement supports the assignment of the low-frequency SdH oscillation in the ST-$CaNbO_3$ film to a small hole-like Fermi pocket associated with the Dirac-like band crossing.

In the relaxed orthorhombic phase, the smallest extremal closed orbit is also hole-like and yields a calculated frequency of approximately 10.2 T, in good agreement with the experimentally observed frequency of 11.53 T for the RO-$CaNbO_3$ film. Additional extremal orbits emerge at higher frequencies, reflecting the more complex multiband Fermi surface topology of the orthorhombic phase The close correspondence between the measured oscillation frequencies and the calculated small Fermi pockets supports a connection between the observed quantum oscillations and electronic states near the predicted symmetry-protected band degeneracies. Taken together, these results show that oxygen-octahedral rotations provide a means of controlling the symmetry-derived electronic structure of $CaNbO_3$, with the calculations indicating an evolution from Dirac-like Fermi pockets in the strained tetragonal phase to nodal-line-like electronic structures in the relaxed orthorhombic phase.

## Discussion

In summary, we demonstrate that oxygen octahedral rotations provide a robust structural route for stabilizing symmetry-derived Dirac electronic states in $CaNbO_3$ thin films. By systematically comparing strained tetragonal and relaxed orthorhombic phases with distinct octahedral rotation patterns, we show that nonsymmorphic symmetry persists across both structural phases and gives rise to consistent Dirac-related transport signatures. Both films exhibit high carrier mobility and multichannel transport, indicative of unconventional charge transport associated with symmetry-derived electronic states. High-field magnetotransport measurements further reveal pronounced Shubnikov–de Haas oscillations arising from small Fermi pockets and finite Berry-phase contributions. First-principles calculations identify a strain-driven evolution from Dirac-electron Fermi pockets in the tetragonal phase to nodal-line-derived electronic structures in the orthorhombic phase, linking the observed transport behavior directly to the underlying oxygen octahedral symmetry.

We would like to further discuss the role of nonsymmorphic symmetry in correlated oxides. Since the early prediction of topological states in pyrochlore iridates[5], studies of topological electronic states in transition-metal oxides have predominantly focused on systems with strong spin–orbit coupling, particularly 5*d* transition-metal oxides, and on geometrically frustrated or triangular-lattice systems. Experimentally realizing such systems in thin-film form can be challenging, as accessing triangular-lattice geometries often requires growth along the (111) direction[48-52], where high-energy surfaces and narrow growth windows can complicate epitaxial stabilization. Direct characterization of their electronic structures using angle-resolved photoemission spectroscopy (ARPES) can also be difficult. On the other hand, recent studies [17-19,24,26,27] have highlighted nonsymmorphic crystal symmetries associated with oxygen-octahedral rotations as an alternative route for stabilizing symmetry-protected electronic states in perovskite oxides, where well-defined epitaxial films and ARPES measurements can provide direct access to the underlying band structure [27]. Thus, our findings establish oxygen-octahedral symmetry as a powerful materials-design parameter for exploring emergent topological electronic phenomena in correlated oxides.

## Methods

**Synthesis of epitaxial $CaNbO_3$ films.**

Epitaxial $CaNbO_3$ films were grown on (001)-oriented $SrTiO_3$ substrates by PLD. Polycrystalline $Ca_2Nb_2O_7$ targets were synthesized by homogeneously mixing dry-milled $CaCO_3$ (99.95%, Sigma Aldrich) and $Nb_2O_5$ (99.99%, Sigma Aldrich), followed by sintering at 1100 °C for 10 h. Prior to deposition, (001) $SrTiO_3$ single crystal substrates (CrysTec GmbH) were chemically etched in buffered oxide etchant and subsequently annealed at approximately 1100 °C to obtain atomically flat $TiO_2$-terminated step-terrace surfaces. The substrates were loaded into the PLD chamber and evacuated to a base pressure of ~ $2 \times 10^{-7}$ Torr. Film deposition was carried out using a KrF excimer laser (Coherent LPXpro 305F, $\lambda = 248$ nm) with a laser fluence of 0.5 J/cm$^2$ and a repetition rate of 5 Hz. High-quality $CaNbO_3$ films were obtained under optimized growth conditions of $T_g = 650$ ºC and $P(O_2) = 4 \times 10^{-6}$ Torr. After deposition, the samples were cooled to room temperature at the rate of 30 °C/min.

**Structural measurements**

X-ray diffraction measurements were performed using a high-resolution four-circle X-ray diffractometer (X'Pert Pro, PANalytical) employing Cu $K\alpha_1$ radiation. Cross-sectional transmission electron microscopy (TEM) specimens were prepared using focused ion beam milling (Thermofisher Helios G4), followed by low-energy milling using a Fischione Nanomill system.

**Transport measurements and analyses**

Electrical and quantum transport measurements were performed using a 9 T Physical Property Measurement System (Quantum Design) and a 30 T bitter magnet equipped with a He$^3$ cryostat at the National High Magnetic Field Laboratory (Tallahassee, USA). Consistent and reproducible results were obtained across multiple samples and measurement platforms. Magnetotransport measurements were carried out in van der Pauw geometry using square-shaped samples ($5 \times 5$ $mm^2$) with indium Ohmic contacts ($< 0.5 \times 0.5$ $mm^2$) placed at the sample corners.

**Semiclassical two-band model for explaining nonlinear Hall resistivity**

The nonlinear Hall response was analyzed using a semiclassical two-band transport model within the framework of Boltzmann transport theory. The Hall resistivity is described by:

$$\rho_{xy} = \frac{[(\mu_1^2\, n_1 + \mu_2^2\, n_2) \; + \; (\mu_1\mu_2 \mathrm{B})^2 \; (n_1 + n_2)]}{e\; [(\mu_1|n_1| + \mu_2|n_2|)^2 + \; (\mu_1\mu_2 \mathrm{B})^2 \; (n_1 + n_2)^2]} \tag{1}$$

together with the zero-field longitudinal resistivity condition:

$$\rho_{xx}(\mathrm{B} \; = \; 0) \; = \frac{1}{e(n_1\mu_1 + \; n_2\mu_2)} \tag{2}$$

Here, $n_1$ ($n_2$), and $\mu_1$ ($\mu_2$) denote the carrier density and mobility, respectively, for electron (hole) type charge carriers.

**Theoretical calculation**

Density functional theory (DFT) calculations were performed using the Vienna ab initio simulation package (VASP)[53],[54] with the projector augmented-wave (PAW) method[55]. The PAW potentials constructed within the Perdew–Burke–Ernzerhof (PBE) generalized gradient approximation were used, and the exchange–correlation energy was treated using the PBE functional[56]. Static electronic-structure calculations were carried out for $CaNbO_3$ structural models with tetragonal $a^0a^0c^-$ and orthorhombic $a^+b^-b^-$ oxygen octahedral rotation patterns, corresponding to the strained and relaxed films, respectively. Spin–orbit coupling was included self-consistently to capture the symmetry-protected band degeneracies near the Fermi level. The plane-wave energy cutoff was set to 500 eV, and the electronic convergence criterion was $10^{-5}$ eV. A $\Gamma$-centered $22 \times 22 \times 26$ k-point mesh was used for the electronic-structure and Fermi-surface calculations. Fermi smearing with a broadening parameter of 0.1 eV was employed.

The symmetry properties of the electronic structures were analyzed along the high-symmetry lines of the corresponding Brillouin zones. The strained tetragonal phase was examined to identify nonsymmorphic-symmetry-protected Dirac-like crossings near the Brillouin-zone boundary, whereas the relaxed orthorhombic phase was analyzed to identify nodal-line-like degeneracies stabilized by combined nonsymmorphic and mirror symmetries.

For comparison with quantum oscillation measurements, three-dimensional Fermi surfaces were generated. Extremal orbit analysis was performed for B||c, where the cyclotron motion occurs in the $k_x$–$k_y$ plane. Consecutive $k_x$–$k_y$ slices were examined along $k_z$, and closed Fermi contours were identified for each slice. The enclosed contour area A was converted to the quantum oscillation frequency using the Onsager relation, $F=\hbar A/2\pi e$. Hole-like and electron-like pockets were distinguished from the band dispersion around the Fermi level, and the resulting extremal orbits were compared with the experimentally observed Shubnikov–de Haas frequencies.

## Data availability

The data that support the findings of this study are available from the corresponding authors upon reasonable request.

## Acknowledgements

This work was primarily supported by the U.S. Department of Energy (DOE), Office of Science, Basic Energy Sciences, Materials Science and Engineering Division. The high-magnetic-field measurements were performed at the National High Magnetic Field Laboratory, which is supported by National Science Foundation cooperative agreement no. DMR-1644779, the state of Florida and DOE. This research used resources of the Advanced Photon Source, a DOE Office of Science User Facility, operated for the DOE Office of Science by Argonne National Laboratory under Contract No. DE-AC02-06CH11357. M.K., S.K., and J. M. O. acknowledge the National Research Foundation of Korea (NRF) grant funded by the Korea Government (MSIT) (RS-2025-2545311).

## Author Contributions

H. N. L. and Y. P., conceived the idea and designed the study; Y. P. and J. S. performed the film growth and characterization with the help of S. L. and S. Y.; S. H. K. carried out the DFT calculation. S. O. provided theoretical insights and critical inputs. Y. P., M. K., and S. K. performed high magnetic field measurements with the help of J. M. O., E. K., and K. R. H. Z. conducted and supported experiments for Bragg rod characterization. H.Z. conducted synchrotron X-ray measurements. Y. P., S. H. K., J. S., and H. N. L wrote the manuscript with contributions from all authors.

## Competing financial interests

The authors declare no competing financial interest.

## Data availability

The data that support the findings of this study are available from the corresponding authors upon reasonable request.

## Appendix A. Supplementary data

Supplementary material related to this article can be found online

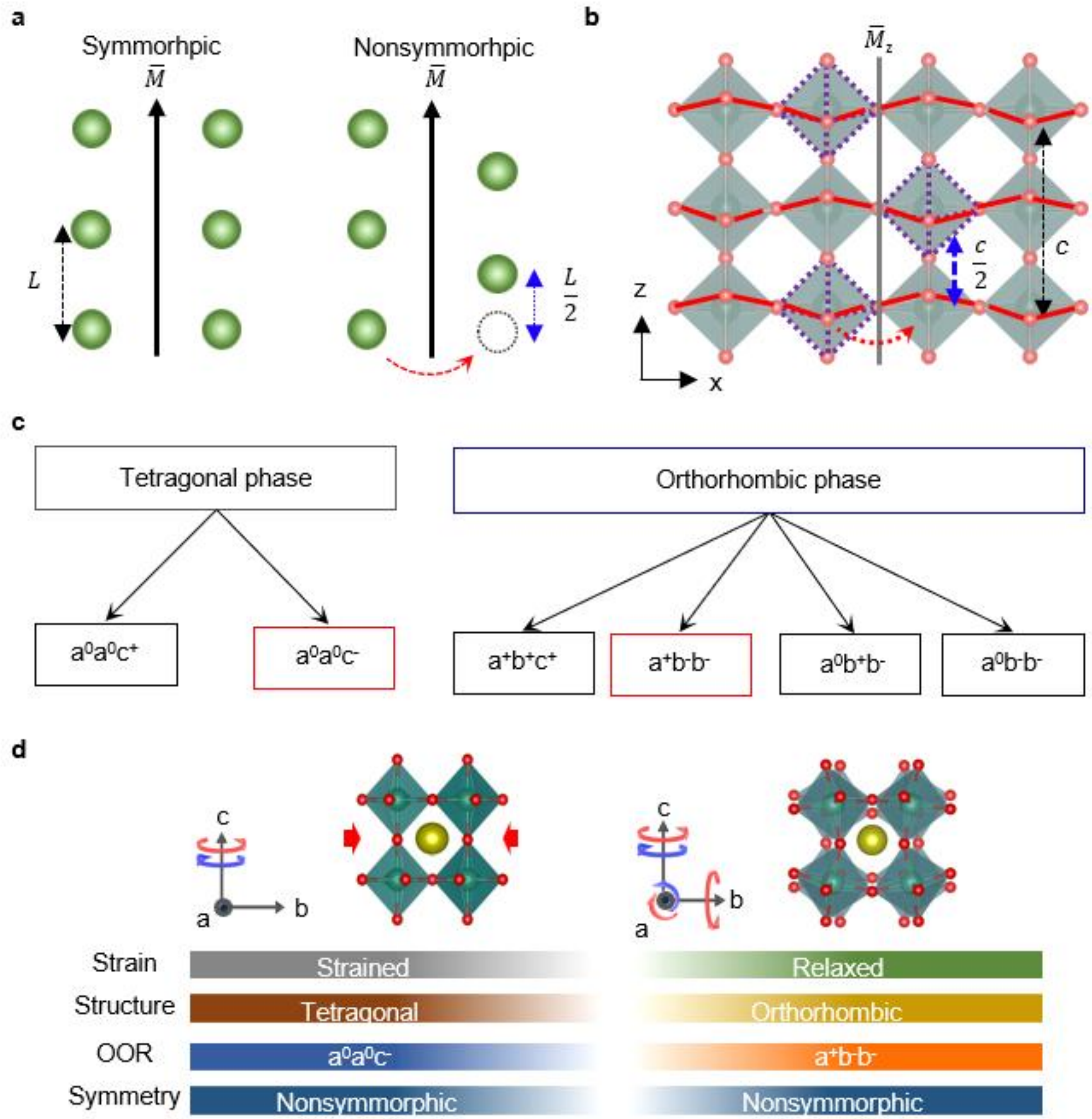

**Fig. 1.** Design principle for strain-controlled nonsymmorphic symmetry in $CaNbO_3$ films. (a) Schematic comparison of symmorphic and nonsymmorphic crystal symmetry. In symmorphic symmetry, all generating symmetry operations leave at least one common point invariant, whereas nonsymmorphic symmetry combines a point-group operation with a fractional lattice translation. (b) Conceptual illustration of nonsymmorphic symmetry in perovskite oxides mediated by oxygen octahedral rotations (OOR). The highlighted octahedra undergo mirror reflection (red dashed arrow) combined with half-unit-cell translation (blue dashed arrow), forming a glide-symmetry operation characteristic of nonsymmorphic structures. (c) Representative OOR patterns in tetragonal and orthorhombic perovskites that preserve nonsymmorphic symmetry (red box). The tetragonal $a^0a^0c^-$ rotation pattern and the orthorhombic $a^0b^+b^-$ and $a^+b^-b^-$ patterns preserve nonsymmorphic symmetry. (d) Strain-controlled OOR with nonsymmorphic symmetry in $CaNbO_3$

films. Under compressive strain (left), $CaNbO_3$ stabilizes a tetragonal structure with $a^0a^0c^-$ (space group: *I4*/*mcm*), whereas relaxed $CaNbO_3$ (right) adopts an orthorhombic $a^+b^-b^-$ structure (space group: *Pnma*).[28]

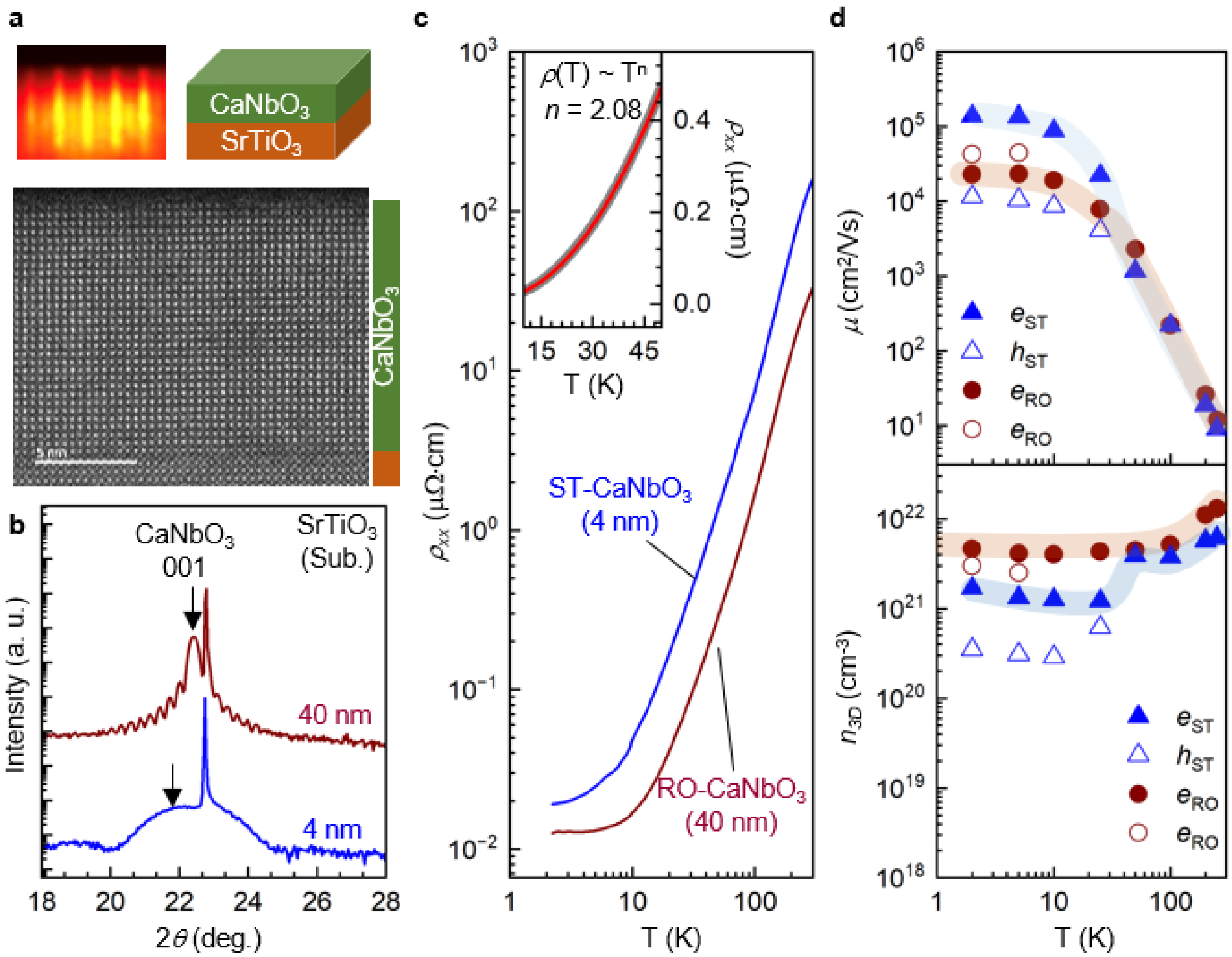

**Fig. 2.** Structural and transport properties of oxygen-octahedral-symmetry-controlled $CaNbO_3$ films. (a) Reflection high-energy electron diffraction (RHEED) pattern (top left), schematic illustration of the epitaxial strain control in $CaNbO_3/SrTiO_3$ heterostructure (top right), and cross-sectional high-angle annular dark-field scanning transmission electron microscopy (HAADF-STEM) image (bottom) of the epitaxial $CaNbO_3$ film on $SrTiO_3$. The STEM image reveals atomically sharp interfaces and coherent epitaxial growth. (b) X-ray diffraction scans of strained (4 nm) and relaxed (40 nm) $CaNbO_3$ films. Film diffraction peaks are marked by arrows, while the asterisk denotes the substrate peak. (c) Temperature-dependent longitudinal resistivity ($\rho_{xx}$) of strained tetragonal (ST) and relaxed orthorhombic (RO) films. Both films exhibit highly metallic behavior. The inset shows a power-law fit ($\rho \sim T^n$) with $n \approx 2.08$, consistent with Fermi-liquid transport. (d) Temperature-dependent carrier mobility (top) and carrier density (bottom) extracted from nonlinear Hall measurements using a semiclassical two-carrier model. The strained film exhibits both electron- and hole-like carriers, whereas the relaxed film shows two electron-like carrier channels. The strained film reaches an ultrahigh mobility of $\mu \approx 1.38 \times 10^5$ cm$^2$ V$^{-1}$s$^{-1}$ at low temperature.

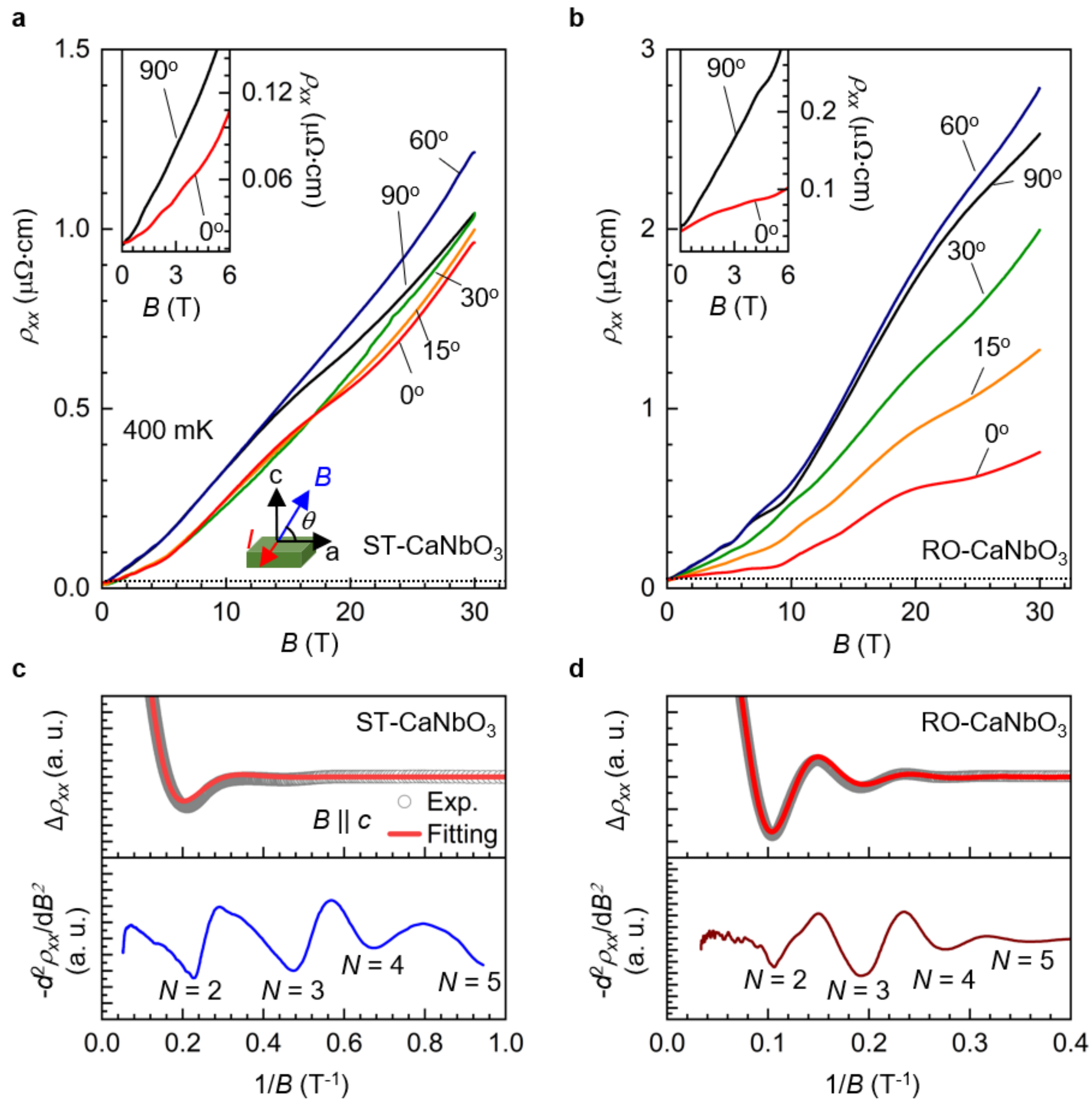

**Fig. 3.** Shubnikov–de Haas (SdH) oscillations and Fermi surface analysis. Magnetic-field-dependent longitudinal resistivity ($\rho_{xx}$) measured at 400 mK for (a) strained tetragonal (ST) and (b) relaxed orthorhombic (RO) $CaNbO_3$ films at various field rotation angles with the magnetic field applied perpendicular to the current (B⊥I). Oscillatory component of the magnetoresistivity, $\Delta\rho_{xx}$, and second derivative of the magnetoresistivity, $-d^2\rho_{xx}/dB^2$ as a function of inverse magnetic field ($1/B$) for (c) ST and (d) RO films. Red lines represent fits using the Lifshitz–Kosevich formalism. Landau level indices ($N$) are assigned to the extrema of $-d^2\rho_{xx}/dB^2$.

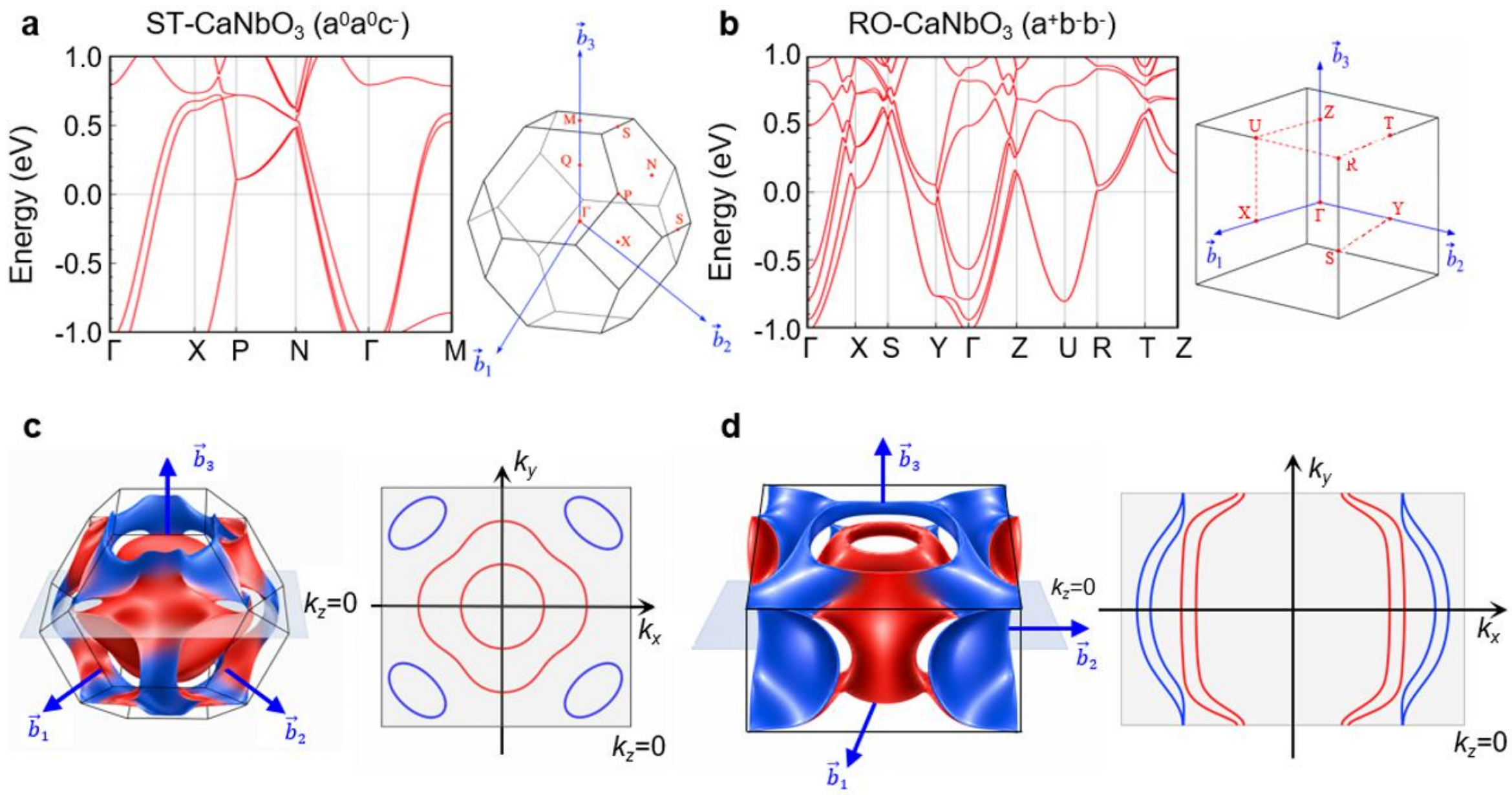

**Fig. 4.** Symmetry-protected electronic states and Fermi surfaces. (a) and (b) Electronic band structures and corresponding Brillouin zones of the strained tetragonal (ST-$CaNbO_3$) $a^0a^0c^-$ and relaxed orthorhombic (RO-$CaNbO_3$) $a^+b^-b^-$ phases, respectively. The strained phase hosts Dirac band crossings near the Fermi level protected by nonsymmorphic symmetry associated with the $a^0a^0c^-$ OOR pattern. In contrast, the relaxed orthorhombic phase exhibits nodal-line-derived degeneracies stabilized by combined nonsymmorphic and mirror symmetries as highlighted by the red dashed lines. (c) and (d) Three-dimensional Fermi surfaces and representative $k_x$–$k_y$ cross-sectional contours at $k_z = 0$ for B||z in the strained and relaxed orthorhombic phases, respectively. Blue and red contours denote hole-like and electron-like Fermi pockets, respectively. The strained phase shows small closed Fermi pockets associated with isolated Dirac crossings, whereas the relaxed orthorhombic phase displays more anisotropic multiband Fermi surfaces originating from nodal-line-like band features. These symmetry-derived Fermi pockets provide a microscopic origin for the low-frequency quantum oscillations observed in both $CaNbO_3$ films.